\documentstyle[12pt]{article}
\newcommand{\thspace}{\kern.08333em}

\def \b{{\cal B}}

\def \beqn{\begin{eqnarray}}
\def \beq{\begin{equation}}

\def \cn{Collaboration}

\def \eeq{\end{equation}}
\def \eeqn{\end{eqnarray}}

\def \s{\sqrt{2}}

\begin{document}

\rightline{SLAC-PUB-7945}
\rightline{EFI-98-46}
\rightline{hep-ph/9809384}

\bigskip
\bigskip
\begin{center}
\Large\bf \boldmath Combining CP Asymmetries in $B\to K \pi$ Decays
\unboldmath
\end{center}

\bigskip
\centerline{Michael Gronau\footnote{Permanent Address: Physics
Department,
Technion -- Israel Institute of Technology, 32000 Haifa, Israel.}}
\centerline{\it Stanford Linear Accelerator Center}
\centerline{\it Stanford University, Stanford, CA 94309}
\medskip
\centerline {and}
\medskip
\centerline{Jonathan L. Rosner}
\centerline{\it Enrico Fermi Institute and Department of Physics}
\centerline{\it University of Chicago, Chicago, IL 60637}
\vskip 1cm

\centerline{\bf ABSTRACT}
\bigskip

\begin{quote}
We prove an approximate relation, to leading order in dominant terms, between
CP-violating rate differences in $B^0/\overline{B}^0 \to K^{\pm}\pi^{\mp}$ and
$B^{\pm} \to K^{\pm}\pi^0$. We show how data from these two processes may be
combined in order to enhance the significance of a nonzero result.
\end{quote}
\bigskip

\leftline{\qquad PACS codes:  12.15.Hh, 12.15.Ji, 13.25.Hw, 14.40.Nd}
\bigskip

Up to now, CP violation has only been observed in the mixing of neutral $K$
meson states \cite{CP64}. Thus, it remains to be confirmed that CP violation in
the kaon system arises from phases in the Cabibbo-Kobayashi-Maskawa matrix
\cite{CKM} describing weak charge-changing transitions of quarks. Such evidence
can be provided by $B$ meson decays, in which the standard model predicts
sizable CP asymmetries between partial rates of $B$ mesons and their
corresponding antiparticles \cite{REVCPB}. Model-dependent calculations of CP
asymmetries in $B$ decays to a pair of charmless pseudoscalar mesons have been
carried out by a large number of authors \cite{MODEL}.

In the present Letter we study relations between direct CP asymmetries in $B
\to K\pi$ decays, following from a model-independent hierarchy among various
contributions to decay amplitudes. Observation of three of these decays, $B^0
\to K^+\pi^-,~B^+\to K^+\pi^0$ and $B^+ \to K^0\pi^+$, combining processes with
their charge conjugates, was reported recently by CLEO \cite{CLEOold, CLEOnew}.
The number of events in these modes is 43, 38, and 12, respectively.
We shall show in this note that while each individual measurement is unlikely
to provide a significant nonzero asymmetry measurement at present levels of
statistics, the rate differences in the first two processes are expected to be
related:
\beq
\Gamma(B^0 \to K^+ \pi^-) - \Gamma(\bar B^0 \to K^- \pi^+) \simeq
2[\Gamma(B^+ \to K^+ \pi^0) - \Gamma(B^- \to K^- \pi^0)]~~~.
\eeq                                                   |
With present statistics and with an estimate of the maximum asymmetry (46\%)
possible in the standard model, the combined sample of $K^\pm \pi^\mp$ and
$K^\pm \pi^0$ events is sufficiently large to display a suitably averaged
asymmetry of up to four standard deviations.
A somewhat more conservative estimate follows from considering rate differences.

In order to study $B \to K\pi$ decays, we will employ a diagrammatic approach
based on flavor SU(3) \cite{GHLR}. Since we concentrate on strangeness-changing
processes, the major part of our analysis will only require isospin symmetry.
SU(3) symmetry and SU(3) breaking effects \cite{SU3BR} will be introduced when
relating these processes to corresponding strangeness-conserving $B \to\pi\pi$
decays. The decomposition of decay amplitudes in terms of flavor flow
topologies is \cite{EWP}
\beqn
-A(B^0 \to K^+\pi^-) &=& (P + {2\over 3}P^c_{EW}) + (T) = B_{1/2} - A_{1/2}
- A_{3/2}~,\cr
-\s A(B^+ \to K^+\pi^0) &=& (P + P_{EW} + {2\over 3}P^c_{EW}) +(T + C +
A) = B_{1/2} + A_{1/2} - 2A_{3/2}~,\cr
A(B^+ \to K^0\pi^+) &=& (P - {1\over 3}P^c_{EW}) + (A)  = B_{1/2} + A_{1/2}
+ A_{3/2}~, \cr
\s A(B^0 \to K^0\pi^0) &=& (P - P_{EW} -{1\over 3}P^c_{EW}) -(C)
= B_{1/2} - A_{1/2} + 2A_{3/2}~.
\eeqn
On the right-hand-sides we also include an equivalent decomposition in terms of
isospin amplitudes \cite{KPIiso}, where $A$ and $B$ are $\Delta I =1$ and
$\Delta I =0$ amplitudes and subscripts denote the isospin of $K\pi$. This
equivalence is implied by the relations
\beqn
B_{1/2} &=& (P + {1\over 6}P^c_{EW}) + {1\over 2}(T + A)~,\cr
A_{1/2} &=& ({1\over 3}P_{EW} -{1\over 6}P^c_{EW}) +(-{1\over 6}T +
{1\over 3}C + {1\over 2}A)~,\cr
A_{3/2} &=& -{1\over 3}(P_{EW} + P^c_{EW}) - {1\over 3}(T + C)~.
\eeqn

The terms in the first parenthesis of Eqs.~(1) and (2), a QCD penguin ($P$), an
electroweak penguin ($P_{EW}$) and a color-suppressed electroweak penguin
($P^c_{EW}$) amplitude, carry each a weak phase ${\rm Arg}(V^*_{tb}V_{ts}) =
-\pi$. The other three terms, tree ($T$), color-suppressed ($C$) and
annihilation ($A$) amplitudes, carry a different weak phase ${\rm
Arg}(V^*_{ub}V_{us})= \gamma$.

We will assume a hierarchy among amplitudes carrying the same weak phase
\cite{EWP}
\beq
\vert P \vert \gg \vert P_{EW} \vert \gg \vert P^c_{EW} \vert~,
\eeq
\beq
\vert T \vert \gg \vert C \vert \gg \vert A \vert~,
\eeq
where a roughly common hierarchy factor of about $0.2$ describes the ratio of
sequential amplitudes. The hierarchy between penguin amplitudes is based on QCD
and electroweak loop factors and is supported by model calculations of short
distance operator matrix elements \cite{FLEI}. Very recently this hierarchy was
shown to follow from model-independent considerations \cite{GPY}. The hierarchy
between $C$ and $T$ is taken from short distance QCD corrections and
phenomenology of $B \to \overline{D}\pi$ decays \cite{Neubert}. The measured
rates of color-suppressed processes, such as $B^0\to\overline{D}^0\pi^0$
\cite{CLEOcolor}, show that rescattering effects do not enhance $C$ to the
level of $T$. This will also be assumed to be the case for $B \to K \pi$.
Finally, the hierarchy between $A$ and $C$ follows essentially from a $f_B/m_B$
factor in $A$ relative to $T$ \cite{Xing}. Several authors \cite{GR,Rescat}
have noted recently that the last assumption, $\vert A \vert \ll \vert C
\vert$, can be spoiled by rescattering effects (from intermediate states
mediated by $T$) through soft annihilation or up-quark penguin topologies. We
will therefore leave open the possibility that $\vert A \vert \sim \vert C
\vert$.
The possibility that $|A|$ can be as large as $|T|$, implied by some
model-dependent calculations \cite{Desh}, will be excluded.  We consider it
unlikely in view of existing limits on rescattering in $B^0 \to K^+ K^-$
\cite{GR}.

Interference between amplitudes carrying different weak phases and different
strong phases leads to CP rate differences between the processes in Eqs. (1)
and their charge conjugates. Such interference involves the product of the
magnitudes of the amplitudes appearing in the first parenthesis with the
amplitudes in the second parenthesis, a sine factor of their relative weak
phase and a sine of the relative strong phase. Thus, all the contributions are
proportional to $\sin\gamma$, whereas the strong phase difference is generally
unknown and may depend on the product. We denote by $2\vec{P}\vec{T}$ the
interference between $P$ and $T$ contributing to $\Delta(K^+\pi^-)\equiv
\Gamma(B^0\to K^+\pi^-) - \Gamma(\overline{B}^0 \to K^-\pi^+)$, and use similar
notations for other interference terms and other CP rate differences. One then
finds for the $B$-$\overline{B}$ rate differences the following expressions,
where terms are written in decreasing order using Eqs.~(3) and (4), and the
smallest terms are neglected:
\beqn
\Delta(K^+\pi^-) &=& 2\vec{P}\vec{T} + {4\over 3}\vec{P}^c_{EW}\vec{T}~,\cr
\Delta(K^+\pi^0) &=& \vec{P}\vec{T} +\vec{P}_{EW}\vec{T} + \vec{P}\vec{C}
+ \vec{P}\vec{A}+ \vec{P}_{EW}\vec{C} + {2\over 3}\vec{P}^c_{EW}\vec{T} +
...~,\cr
\Delta(K^0\pi^+) &=& 2\vec{P}\vec{A} +...~,\cr
\Delta(K^0\pi^0) &=& -\vec{P}\vec{C} + \vec{P}_{EW}\vec{C} +
{1\over 3}\vec{P}^c_{EW}\vec{C}~.
\eeqn
We note that, in the absence of electroweak penguin amplitudes, one finds
\cite{SONI}
\beq \label{eqn:AS}
\Delta(K^+\pi^-) + \Delta(K^0\pi^+) = 2\Delta(K^+\pi^0) + 2\Delta(K^0\pi^0)~.
\eeq
However, this relation is spoiled by electroweak penguin contributions.

Comparing the four rate differences, we see that the dominant terms of the form
$\vec{P}\vec{T}$ appear only in the first two rate differences, leading at this
order to the relation
\beq\label{Rel}
\Delta(K^+\pi^-) \approx 2\Delta(K^+ \pi^0)~.
\eeq
The next-to-leading terms correcting this relation are $\vec{P}_{EW}\vec{T}$
and $\vec{P}\vec{C}$. The first term can be shown to lead to a negligible rate
difference. The argument is based on a property of the $A_{3/2}$ amplitude,
which was shown recently \cite{NR} to consist of $T + C$ and electroweak
penguin contributions with approximately equal strong phases. Using this
property we conclude that
\beq
\vec{P}_{EW}\vec{T} + \vec{P}_{EW}\vec{C} + \vec{P}^c_{EW}\vec{T} +
\vec{P}^c_{EW}\vec{C} \approx 0~,
\eeq
or, to leading order, that $\vec{P}_{EW}\vec{T}\approx 0$. Since the term
$\vec{P}\vec{C}$ is the only next-to-leading correction to Eq.~(\ref{Rel}),
this equality is expected to hold to about $20\%$.

Using the hierarchy $\vert A\vert \ll \vert C\vert$, it has often been assumed
that the rate difference $\Delta(K^0\pi^+)$ is extremely small. However,
recently it was argued \cite{GR,Rescat} that rescattering effects may enhance
$A$ to the level of $C$, thus leading to a CP asymmetry in this process at a
level of $10\%$. This would imply that the term $\vec{P}\vec{A}$ appearing in
both $\Delta(K^0\pi^+)$ and $\Delta(K^+\pi^0)$ is next-to-leading and may be
comparable to $\vec{P}\vec{C}$. In this case Eq.~(\ref{Rel}) could be violated
by up to about 40\%, and a better approximation becomes
\beq \label{eqn:approx}
\Delta(K^+\pi^-) + \Delta(K^0\pi^+) \approx 2\Delta(K^+ \pi^0)~.
\eeq
A way of gauging the importance of the $\vec{P}\vec{A}$ term would be by
measuring a nonzero value for $\Delta(K^0\pi^+)$. The dominant correction to
the approximate relation (\ref{eqn:approx}) is the term $-2 \vec{P} \vec{C}$
which is contributed by $2 \Delta(K^0 \pi^0)$ on the right-hand side of
(\ref{eqn:AS}).

In order to study relative asymmetries, let us first discuss the rates
themselves. To leading order, all four $B \to K \pi$ processes are dominated by
the (gluonic) penguin terms $P$ in Eq.~(1), and their rates as well as their
charge-conjugates rates are expected to satisfy the relations
\beq\label{rates}
\Gamma(B^+ \to K^0 \pi^+) = \Gamma(B^0 \to K^+ \pi^-)
= 2 \Gamma(B^+ \to K^+ \pi^0) = 2 \Gamma(B^0 \to K^0 \pi^0)
\eeq
The next-to-leading corrections to these equalities are interference terms of
the form $2{\rm Re}(PT^*)$ and $2{\rm Re}(PP^*_{EW})$. To first order in small
quantities, the $B \to K \pi$ rates satisfy the sum rule \cite{Lipkin}
\beq \label{eqn:sumrule}
2 \Gamma(B^+ \to K^+ \pi^0) + 2 \Gamma(B^0 \to K^0 \pi^0)
= \Gamma(B^+ \to K^0 \pi^+) + \Gamma(B^0 \to K^+ \pi^-)~~~,
\eeq
which may be used to anticipate a small $B^0 \to K^0 \pi^0$ rate.

The first order terms modify the rate equalities Eq.~(\ref{rates}). The
difference between $2\b(B^+ \to K^+ \pi^0)$ and $\b(B^+ \to K^0 \pi^+)$ is
given by two terms, $2{\rm Re}(PT^*) + 2{\rm Re}(PP^*_{EW})$, which can imply a
factor as large as about two between these branching ratios. In fact, the
present central values measured for these branching ratios differ by a factor
of 2.1 \cite{CLEOnew}, which can be used \cite{NR} to place model-independent
bounds on the weak phase $\gamma$. The rate difference of the two processes
involving charged kaons, to which Eq.~(\ref{Rel}) applies, is given to this
order by \beq 2\Gamma(B^+\to K^+ \pi^0) - \Gamma(B^0 \to K^+\pi^-) = 2{\rm
Re}(PP^*_{EW})~, \eeq where the correction term is expected to be no larger
than about 40\% of each of these rates. Since a similar approximation applies
to the rate difference (or a better one in case of small rescattering effects),
the asymmetries of these two processes can differ by about a factor of two. In
general, with no other information about the above interference terms, the
equality between rate differences Eq.~(\ref{Rel}) holds to a better
approximation than the equality of corresponding asymmetries.

In order to estimate the CP asymmetries in $B^0/\overline{B}^0 \to
K^{\pm}\pi^{\mp}$ and $B^{\pm} \to K^{\pm}\pi^0$, one must know the
ratio $\vert T/P\vert$. Using previous data \cite{CLEOold} we have shown
\cite{DGR} that this ratio is smaller than one, representing another hierarchy
factor of about 0.2. Let us update information about $P$ and $T$ using more
recent data \cite{CLEOnew}.
We will quote squares of amplitudes in terms of decay rates.

The most straightforward way of obtaining $\vert P\vert$ is from the observed
CP-averaged branching ratio \cite{CLEOnew}
\beq
\b(B^{\pm} \to K^0/\overline{K}^0 \pi^{\pm}) = (14 \pm 5 \pm 2)\times
10^{-6}~,
\eeq
since there are no first order corrections to $P$ in these processes even when
$\vert A\vert$ is as large as $ \vert C \vert$.
Using the value \cite{CLEOnew} $\tau(B^+) = (1.65 \pm 0.04) \times
10^{-12}{\rm~s}$, we find $|P|^2 = \Gamma(B^\pm \to K^0/\bar K^0 \pi^\pm) =
(8.5 \pm 3.3) \times 10^6 {\rm~s}^{-1}$.

An estimate of $\vert T\vert$ is more uncertain at this time. While CLEO has
quoted upper limits at $90\%$ confidence level:
\beq \label{LIM+0}
\b(B^0/\overline{B}^0 \to \pi^+\pi^-) < 8.4\times 10^{-6}~~,~~~
\b(B^{\pm} \to \pi^{\pm}\pi^0) < 16\times 10^{-6}~,
\eeq
their data imply signals for these decays with significance of 2.9 and 2.3
standard deviations, respectively.  Taking these signals seriously, we may
obtain from the reported event rates and efficiencies the branching ratios
\beq
\b(B^0/\overline{B}^0 \to \pi^+ \pi^-) = (3.7_{-1.7}^{+2.0})\times 10^{-6}~,
\eeq
\beq
\b(B^{\pm} \to \pi^{\pm} \pi^0) = (5.9_{-2.7}^{+3.2})\times 10^{-6}~.
\eeq
While destructive interference between tree and penguin amplitudes in $B^0\to
\pi^+\pi^-$ and/or constructive interference between tree and color-suppressed
or electroweak penguin amplitudes in $B^+ \to \pi^+ \pi^0$ may lead to $\b(B^0
\to \pi^+\pi^-) < 2\b(B^+\to\pi^+\pi^0)$ \cite{GHLR}, we shall ignore such
effects as in Ref.~\cite{DGR}.  Thus, using an SU(3) relation between $B\to K
\pi$ and $B\to \pi\pi$, taking $\tau(B^0) = (1.56 \pm 0.04) \times 10^{-12}
{\rm~s}^{-1}$, and introducing SU(3) breaking through $f_K/f_{\pi}$
\cite{GHLR,SU3BR}, we have the two independent estimates
\beq
|T|^2_{B^0} = \left[ {V_{us}\over V_{ud}}{f_K\over f_{\pi}} \right]^2
\Gamma(B^0/\bar B^0 \to \pi^+\pi^-) = (1.8 \pm 0.9) \times 10^5 {\rm~s}^{-1}~~,
\eeq
\beq
|T|^2_{B^\pm} = 2 \left[ {V_{us}\over V_{ud}}{f_K\over f_{\pi}} \right]^2
\Gamma(B^\pm \to \pi^\pm \pi^0) = (5.4 \pm 2.7) \times 10^5 {\rm~s}^{-1}~~~,
\eeq
whose average is $|T|^2 = (2.16 \pm 0.85) \times 10^5 {\rm~s}^{-1}$.  When
combined with our estimate for $|P|^2$, and assuming an additional 20\% error
from neglecting a penguin amplitude in $B^0 \to \pi^+ \pi^-$ and a
color-suppressed amplitude in $B^+ \to \pi^+\pi^0$, this leads to $|T/P| =
0.160 \pm 0.054$, or $|T/P| < 0.23$ at 90\% confidence level. A more precise
determination of this ratio requires more statistics. We will assume its
preliminary value. A slightly larger value of $|T+C|/|P| = 0.24 \pm 0.06$ was
estimated in Refs.~\cite{NR} and \cite{NRnext}.

As we have shown, CP asymmetries in $B^0/\overline{B}^0 \to K^{\pm}\pi^{\mp}$
and $B^{\pm} \to K^{\pm}\pi^0$ are equal to each other, to leading order in
$\vert T/ P\vert$, $\vert P_{EW}/P\vert$ and $\vert C/T\vert$, and are given by
$2\vert T/ P\vert\sin\gamma\sin\delta$, where $\delta$ is the strong phase
difference between $T$ and $P$. This phase is generally unknown, but could be
substantial. While the tree amplitudes is expected to factorize, thus showing
little evidence for rescattering effects, the penguin amplitude obtains a large
contribution from a so-called charming penguin term \cite{CHARMPEN}, involving
long distance effects of rescattering from charm-anticharm intermediate states.
It is therefore conceivable that $\delta$ could attain a large value, such that
$\sin\delta\sim 1$. The values of $\gamma$ allowed at present \cite{JR} include
those around $90^{\circ}$ obeying $\sin\gamma\sim 1$. We therefore conclude
that an interesting range of asymmetry measurements includes the value $2\vert
T/P\vert$ which we found to be $0.32 \pm 0.11$ to leading order, or less than
46\% at 90\% confidence level.

To first approximation, CP asymmetries in the processes $B^0/\overline{B}^0 \to
K^\pm \pi^\mp$ and $B^\pm \to K^\pm \pi^0$ are equal. Averaging them leads to a
statistically more significant result then measuring them separately. Denoting
by $N_n,~A_n$ and $N_c,~A_c$ the number of events and asymmetries in
$B^0/\overline{B}^0 \to K^\pm \pi^\mp$ and $B^\pm \to K^\pm \pi^0$,
respectively, one may define an averaged asymmetry
\beq
A_{\rm av} \equiv \frac{N_n A_n + N_c A_c}{N_{\rm tot}}~~~,
\eeq
where $N_{\rm tot}$ is the total number of $K^\pm \pi^0$ and $K^\pm \pi^\mp$
events. It is easy to show that, under the assumption of equal asymmetries, the
total number of events $N_{\rm tot}$ required to observe this asymmetry at the
$n$-standard-deviation level of significance does not exceed $N_{\rm tot} =
(n/A_{\rm av})^2$.  Thus, for $|A_{\rm max}| = 0.46$ and $N_{\rm tot} = 43 + 38
= 81$ events, one could see a signal as large as four standard deviations,
whereas the maximum signals based on $N_n$ and $N_c$ separately would not be
expected to exceed $3 \sigma$. Backgrounds
and particle misidentification will degrade these estimates somewhat.

As mentioned, in general the approximate equality of CP rate-differences in the
processes $B^0/\overline{B}^0 \to K^\pm \pi^\mp$ and $B^\pm \to K^\pm \pi^0$ is
expected to hold to a better accuracy than the equality of corresponding
asymmetries.  Thus, we shall estimate the errors on the separate quantities
\beq
(\vec{P} \vec{T})_n \equiv \Delta(K^+ \pi^-)/2~~,~~~
(\vec{P} \vec{T})_c \equiv \Delta(K^+ \pi^0)
\eeq
and on their average, and compare these with the maximum possible value of
\beq
\vec{P} \vec{T} = 2|P||T| \sin(\delta_T - \delta_P) \sin \gamma~~~.
\eeq

The most conservative estimate is based directly on the experimental errors on
sums of rates for particles and antiparticles, which one may show are equal to
the errors on rate {\it differences}: $\delta \Delta(K^+ \pi^-) = \delta
\Gamma_n$, $\delta \Delta(K^+ \pi^0) = \delta \Gamma_c$, where 
\beq
\Gamma_n \equiv \Gamma(B^0 \to K^+ \pi^-) + \Gamma(\bar B^0 \to K^- \pi^+)~~,~~
\Gamma_c \equiv \Gamma(B^+ \to K^+ \pi^0) + \Gamma(B^- \to K^- \pi^0)~~~.
\eeq
(In practice $\delta \Delta(K^+ \pi^-)$ may exceed $\delta \Gamma_n$ if there
is a kinematic ambiguity between $K^+ \pi^-$ and $K^- \pi^+$ final states.)
Using branching ratios
\beq
(1/2)[{\cal B}(B^0 \to K^+ \pi^-) + {\cal B}(\bar B^0 \to K^- \pi^+)] =
(14 \pm 3 \pm 1) \times 10^{-6}~~~,
\eeq
\beq
(1/2)[{\cal B}(B^+ \to K^+ \pi^0) + {\cal B}(\bar B^- \to K^- \pi^0)] =
(15 \pm 4 \pm 3) \times 10^{-6}~~,~~~
\eeq
and the $B^0$ and $B^+$ lifetimes mentioned above, we find
\beq
\Gamma_n = (17.9 \pm 4.1) \times 10^6{\rm~s}^{-1}~~,~~~
\Gamma_c = (18.2 \pm 6.1) \times 10^6{\rm~s}^{-1}~~~.
\eeq
These rates [which should obey $\Gamma_n/2 = \Gamma_c$ if $P$ were the only
amplitude present, as noted in Eq.~(\ref{rates})] lead to individual errors on
$\vec{P} \vec{T}$ of 
\beq \label{eqn:pess}
\delta (\vec{P} \vec{T})_n = \delta \Gamma_n/2 = 2.0 \times 10^6{\rm~s}^{-1}~~,
~~\delta (\vec{P} \vec{T})_c = \delta \Gamma_c = 6.1 \times 10^6{\rm~s}^{-1}~~,
\eeq
and a combined error of
\beq \label{eqn:err}
\delta (\vec{P} \vec{T})_{\rm comb} = 1.9 \times 10^6{\rm~s}^{-1}~~~.
\eeq

The maximum value of $\vec{P} \vec{T}$ is
\beq \label{eqn:max}
|\vec{P} \vec{T}|_{\rm max} = 2\sqrt{|P|^2 |T|^2} = (2.71 \pm 0.75)
\times 10^6{\rm~s}^{-1}~~~,
\eeq
based on the estimates of $|P|^2$ and $|T|^2$ given above. Comparing the error
(\ref{eqn:err}) with the value (\ref{eqn:max}), one sees that reduction of the
error by a factor of anywhere from about $\sqrt{3}$ to 3 could permit a
non-zero observation of $\vec{P} \vec{T}$ at the $3 \sigma$ level.

In the ideal case in which fractional rate errors scale as $1/\sqrt{N}$,
Eq.~(\ref{eqn:pess}) would be replaced by
\beq \label{eqn:opt}
\delta (\vec{P} \vec{T})_n = 1.36 \times 10^6{\rm~s}^{-1}~~,~~
\delta (\vec{P} \vec{T})_c = 2.95 \times 10^6{\rm~s}^{-1}~~~,
\eeq
leading to a combined error of $\delta (\vec{P} \vec{T})_{\rm comb} = 1.24
\times 10^6{\rm~s}^{-1}$, sufficient to demonstrate a $3 \sigma$ effect if
$\vec{P} \vec{T}$ were at the upper limit of its allowed range.

To conclude, we have shown that to leading order in small quantities it makes
sense to combine the CP-violating rate differences in the decays $B^0 \to K^+
\pi^-$ and $B^+ \to K^+ \pi^0$.  Whereas the identification of the flavor of
charged secondaries in $B^0/\overline{B}^0 \to K^\pm \pi^\mp$ decays requires
good particle
identification in order to avoid a kinematic ambiguity involving $\pi
\leftrightarrow K$ interchange, no such ambiguity afflicts the decays $B^\pm
\to K^\pm \pi^0$.  The averaged rate difference can be large enough in the
standard model that it would be detectable at present levels of sensitivity.

\vspace{0.3cm}
{\it Acknowledgments:\/}
We thank Jim Alexander, Yuval Grossman, Matthias Neubert, Jim Smith, Sheldon
Stone, Bruce Winstein, Mihir Worah and Frank W\"urthwein for useful
discussions. M.G. is grateful to the SLAC Theory Group for its kind
hospitality. Part of this work was done during the Workshop on {\sl
Perturbative and Non-Perturbative Aspects of the Standard Model\/} at St.\
John's College, Santa Fe, July--August 1998.  J. L. R. would like to thank the
organizer Rajan Gupta, as well as the participants of the workshop, for
providing a stimulating atmosphere and for many useful discussions. This work
was supported in part by the United States Department of Energy through
contracts DE FG02 90ER40560 and DE AC03 76SF00515, and by the United States -
Israel Binational Science Foundation under Research Grant Agrement
94-00253/2.
\newpage

\def \ajp#1#2#3{Am.~J.~Phys.~{\bf#1}, #2 (#3)}
\def \apny#1#2#3{Ann.~Phys.~(N.Y.) {\bf#1}, #2 (#3)}
\def \app#1#2#3{Acta Phys.~Polonica {\bf#1}, #2 (#3)}
\def \arnps#1#2#3{Ann.~Rev.~Nucl.~Part.~Sci.~#1 (#2) #3}
\def \cmp#1#2#3{Commun.~Math.~Phys.~{\bf#1}, #2 (#3)}
\def \ib{{\it ibid.}~}
\def \ibj#1#2#3{~{\bf#1}, #2 (#3)}
\def \ijmpa#1#2#3{Int.~J.~Mod.~Phys.~A {\bf#1}, #2 (#3)}
\def \ite{{\it et al.}}
\def \jmp#1#2#3{J.~Math.~Phys.~{\bf#1}, #2 (#3)}
\def \jpg#1#2#3{J.~Phys.~G {\bf#1}, #2 (#3)}
\def \mpla#1#2#3{Mod.~Phys.~Lett.~A {\bf#1}, #2 (#3)}
\def \nc#1#2#3{Nuovo Cim.~{\bf#1}, #2 (#3)}
\def \npb#1#2#3{Nucl.~Phys. B~{\bf #1}, #2 (#3)}
\def \pisma#1#2#3#4{Pis'ma Zh.~Eksp.~Teor.~Fiz.~{\bf#1}, #2 (#3) [JETP
Lett. {\bf#1}, #4 (#3)]}
\def \pl#1#2#3{Phys.~Lett.~{\bf#1}, #2 (#3)}
\def \plb#1#2#3{Phys.~Lett.~B {\bf #1}, #2 (#3)}
\def \pr#1#2#3{Phys.~Rev.~{\bf#1}, #2 (#3)}
\def \pra#1#2#3{Phys.~Rev.~A {\bf#1}, #2 (#3)}
\def \prd#1#2#3{Phys.~Rev.~D {\bf #1}, #2 (#3)}
\def \prl#1#2#3{Phys.~Rev.~Lett.~{\bf #1}, #2 (#3)}
\def \prp#1#2#3{Phys.~Rep.~{\bf#1}, #2 (#3)}
\def \ptp#1#2#3{Prog.~Theor.~Phys.~{\bf#1}, #2 (#3)}
\def \rmp#1#2#3{Rev.~Mod.~Phys.~{\bf#1}, #2 (#3)}
\def \rp#1{~~~~~\ldots\ldots{\rm rp~}{#1}~~~~~}
\def \stone{{\it B Decays}, edited by S. Stone (World Scientific,
Singapore, 1994)}
\def \yaf#1#2#3#4{Yad.~Fiz.~{\bf#1}, #2 (#3) [Sov.~J.~Nucl.~Phys.~{\bf #1},
#4 (#3)]}
\def \zhetf#1#2#3#4#5#6{Zh.~Eksp.~Teor.~Fiz.~{\bf #1}, #2 (#3) [Sov.~Phys.
- JETP {\bf #4}, #5 (#6)]}
\def \zpc#1#2#3{Zeit.~Phys.~C {\bf #1}, #2 (#3)}


\begin{thebibliography}{99}
\bibitem{Lipkin} H. J. Lipkin, hep-ph/9809347. As can be seen from Eqs.~(1),
this relation holds also in the presence of electroweak penguin contributions.
\end{thebibliography}
\end{document}